Bogdan Czejdo, Wiktor B. Daszczuk, Mikołaj Baszun

# USING MACHINE LEARNING TO ENHANCE VEHICLES TRAFFIC IN ATN (PRT) SYSTEMS

*This paper discusses new techniques to enhance Automated Transit Networks (ATN, previously called Personal Rapid Transit - PRT) based on Artificial Intelligence tools. The main direction is improvement of the cooperation of autonomous modules that use negotiation protocols, following the IoT paradigm. One of the goals is to increase ATN system throughput by tuning up autonomous vehicles cooperation. Machine learning (ML) was used to improve algorithms designed by human programmers. We used "existing controls" corresponding to near-optimal solutions and built refinement models to more accurately relate a system's dynamics to its performance. A mechanism that mostly influences ATN performance is Empty Vehicle Management (EVM). The algorithms designed by human programmers was used: calls to empty vehicles for waiting passengers and balancing based on reallocation of empty vehicles to achieve better regularity of their settlement. In this paper we discuss how we can improve these algorithms (and tune them to current conditions) by using ML to tailor individual behavioral policies. Using ML techniques was possible because our algorithm is based on a set of parameters. A number of weights and thresholds could be tuned up to give better decisions on moving empty vehicles across the track.*

## INTRODUCTION

New automatization techniques in urban transport are under development. Many of them concern autonomous vehicles moving in urban traffic [1] [2] [3]. Also, automatization techniques are applied for driver support [4]. In Automated Transit Networks (ATN, also called Personal Rapid Transit - PRT [5]), the movement takes place on a separate track, typically raised above the ground level. This approach liberates the designers from most problems of recognition of the environment. Main targets are keeping up the track [6], routing [7] and empty vehicles management [8].

Some modern ATN systems are decentralized, where decisions are made using simple communication protocols between autonomous modules. For example, a delivery of empty vehicles may follow this principle [8] [9]. The distributed cooperation of autonomous modules on Internet of Things paradigm (IoT [10]), using simple negotiation protocols [11] may be used in ATN station maneuvering, where track segment controllers guide vehicles between charging lot, parking lots and boarding/alighting lots [12] [13]. Such a maneuvering may cause traffic conflicts, if more than one vehicle take part in a change of places. An example of an automatic vehicle guidance system based on cooperating track segment controllers is described in [14]. The other subject of autonomous vehicles cooperation is Empty Vehicles Management, in which vehicles reallocate in ATN, anticipating future needs and compared with current state of the network [15] [16].

The network has some *ridership*, counted as a number of trips that may be executed in a time unit, for a given Origin-Destination matrix and a given number of vehicles. Various methods are used to check the ridership [8] [17] but they give similar results. Our approach is simulation-based, in which there are infinite queues of passengers on every station and therefore every vehicle finishing its trip is taken by a waiting passenger group.

Having real demand, the network performs a number of trips lower than the maximum ridership. We call *throughput* a number of full trips (with passengers) preformed in a time unit in real traffic conditions. In addition to a number of vehicles and an Origin-Destination matrix, a demand is a parameter of the throughput. Empty trips organized in the system allow to increase the throughput. Higher throughput shortens the average passenger waiting time (at a station in a queue).

Machine learning (ML) in some problem domains can generate solutions better than algorithms designed by human programmers [18] [19]. There are also domains where using machine solutions can improve existing hardware/software system. Controlling vehicles traffic in ATN systems is one of these domains. Theoretically it is possible to collect enough data either through a simulation or on-line monitoring system to completely "control" ATN Systems by a trained model. Practically, however, it is much more realistic to use "existing controls" based on near-optimal solutions and build refinement models to more accurately relate a system's dynamics to its performance. In Automated Transit Networks we can control in a relatively robust and reliable fashion as shown [8]. A mechanism that mostly influences ATN performance is Empty Vehicle Management (EVM) [8] [16]. The two algorithms are used: calling empty vehicles for waiting passengers and balancing - reallocation of empty vehicles to achieve better regularity of their settlement [8] (or, on the contrary, irregular allocation of vehicles for special events [20]). The goal of these algorithms is better operation of ATN system, expressed as shortening of passenger waiting time at the cost (possibly smallest) of some number of empty trips performed. We will discuss how we can improve Automated Transit Networks performance on a cooperative transportation task by using ML to adjust individual behavior selection polices from models related to global performance. Unlike some EVM algorithms proposed [15] [21] [22], our algorithm is based on a set of parameters measuring physical features of the net (like distances between stations and their capacity), historical data (past demand used to predict future demand) and current traffic conditions (current vehicles allocation and a quantity of passenger groups waiting in queues at the stations). A number of weights and thresholds applied to these features give the decisions on moving empty vehicles across the track.

Choosing of proper set of parameter values is a difficult process because of large number of parameters and long simulation time of every experiment. Machine learning techniques [23] [24] may be applied to obtain best values of a set of parameters governing EVM. The control system may be automatically learned to react properly in typical situations, in which current demand fits historical data, and in a case of unusual situation when a social event or an accident changes the demand rapidly. Also, learning may be performed during every day operation to adopt the control to changes in number of vehicles available, Origin-Destination matrix structure, physical track changes due to vehicle/track malfunction, conservation conditions, change of maximum velocity allowed (atmospheric conditions, bending of the track, etc.).

There are many enhancements for the vehicle traffic in ATN systems based on machine learning. In this paper we classify the efforts based on area of enhancement, data used and machine learning technique. In this paper we discuss efforts related to applying the following techniques: Supervised Learning, Clustering and Reinforcement Learning. Section 1 presents general analysis of machine learning use in ATN systems. In Section 2, an application of neural networks to implement ATN throughput enhancements is discussed. The conclusions are presented in Section 3.

## 1. MACHINE LEARNING FOR ATN SYSTEMS

One of the enhancement of the current control is to generate adjustments for EVM better performance.

Due to huge collection of parameter values sets, which cannot be searched completely, two modes of learning should be applied. In *training* phase, many parameter sets and many working condition parameters should be applied to obtain best tuning of the algorithm for given net structure. Then, during normal *execution* of the system, algorithm parameters should be changed slightly to learn how to fine-tune the algorithm and to adopt it to changing operation conditions.

Many algorithms for EVM are described in the literature, for example [15] [21] [22]. Most algorithms are focused on the optimal reallocation of empty vehicles, and some of them on delivering empty vehicles for waiting passengers. Both of them may reduce the passenger waiting time significantly. Reallocation algorithms are usually based on past demand estimates and future forecast. All mentioned approaches use a form of central data base in which historical demand and actual positions of empty vehicles are stored.

Our algorithm, described in [8], uses decentralized data on demand (passenger queues), past demand, vehicles staying at the stations and on the move (performing full and empty trips). Any station retrieves data from its neighboring stations, which allows for distributed implementation. If the demand follows historical trends, the algorithm acts like typical predictive ones. Yet, in a case of non-typical situation (for example a social event organized in a given quarter of a town), other parameters cause the change in EVM to cover this unusual demand [20].

In fact, two EVM algorithms with identical structure but separate sets of parameters are used: calling for vehicle delivery and balancing for vehicle redistribution. Both of them cooperate to enhance the network throughput, which is measured by passenger waiting time.

Based on our model we can generate data for machine learning processing.

For the calculation of the values of the two functions (for calling and balancing algorithms), a vector of weighting factors and threshold values has been defined. The weighting factors determine how strongly given parameters influence the decision to move a given vehicle. The thresholds define minimum values of measured features, which allow a vehicle to move.

– $F_Q$ — passenger queue factor — determines the impact of the passenger queue length in target station;
– $F_{EB}$ — empty berths factor — the impact of a number of empty berths in target station or capacitor;
– $F_{ND}$ — normalized inverse distance factor — the impact of normalized inverse distance between nodes ($ND_{ij}=D_{av}/D_{ij}$, Where $D_{ij}$ — shortest distance from station $s_i$ to station $s_j$, $D_{av}$ — average distance between a pair of distinct stations; Notice that the distance between stations $D_{ij}$ is a denominator: the shorter the actual distance is, the greater is the value of $ND_{ij}$; $ND=1$ for mean distance).
- $F_{AI}$ — historical demand factor — the impact of mean value of passenger groups inter-arrival time distribution at target station during previous days (a measure of predicted demand); the mean value is a denominator, because the shorter is the time between occurrences of two consecutive groups, the stronger the impact is;
- $T_Q$ — passenger queue threshold — if in a queue there are less passenger groups than $T_Q$, then a vehicle is not moved;
- $T_{EB}$ — empty berths threshold — if there are less empty berths than $T_{EB}$, then a vehicle is not moved;
- $T_{EV}$ — empty vehicles threshold — if there are less empty vehicles in berths than $T_{EV}$, then a vehicle is not moved;
- $T_{ND}$ — normalized inverse distance threshold (inverse of the horizon) — if the distance between nodes is greater than $T_{ND}$ (note that the actual distance between nodes is a denominator), then a vehicle is not moved;
- $T$ — total function threshold — if the value calculated as the sum of products of individual factors by corresponding static or dynamic parameter values is less than $T$, then a vehicle is not moved.

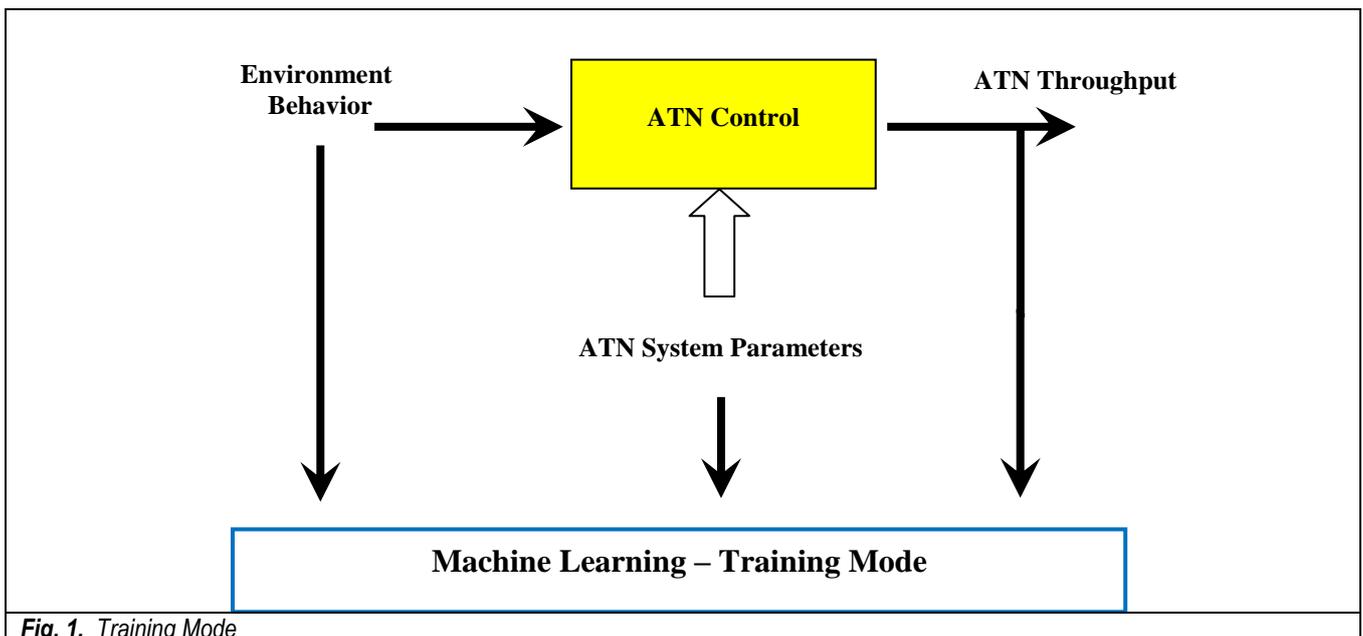

*Fig. 1.* Training Mode

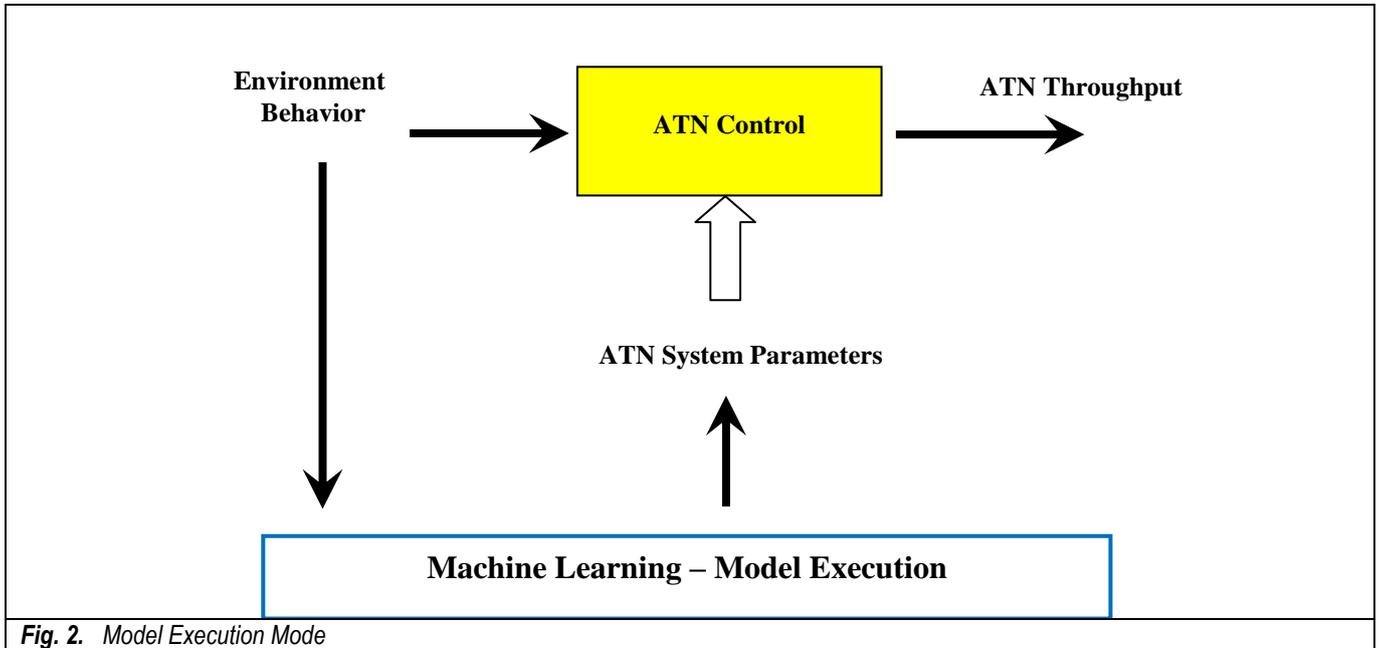

*Fig. 2.* Model Execution Mode

Each function has its separate set of the above weighting factors and thresholds. The factors and thresholds for the balancing function have *B* prefix (i.e., $BF_Q$, $BF_{EB}$, etc.), and for the calling function they have *C* prefix (i.e., $CF_Q$, $CF_{EB}$, etc.). The details of the algorithm may be found in [8].

In addition to the algorithm parameters, some values describe the operating conditions:
- Number of vehicles;
- Maximum velocity allowed;
- Actual demand: total and for selected stations (for example 4);
- Several structures of Origin-Destination matrix (for example 4).

This gives a set of 11 additional parameters.

The simulation methods can provide verification of some solutions satisfying the main constrains but simulation methods alone cannot guarantee an optimal solution in all circumstances. Various approaches have been proposed that heavily sample the search space in search for the global optimal solution but most of them are very expensive and too slow for online applications especially for significantly changing environment requirements. There is no global optimization method that can solve this problem with any theoretical guarantee of success. Combining simulation with machine learning can improve this process in many aspects

New applications of Machine Learning (ML) [25] are identified very rapidly. These changes are facilitated by new approaches and breakthroughs in parallelization of artificial neural networks algorithms that takes advantage of fast and parallel hardware [24]. Each new application, however, requires careful determination of training data. When using Deep Learning approach the ML model can be created from raw data and labels assuming that we have enough labeled data. In a more traditional approach the raw data is first transformed to, so called, feature vector and then the training is performed on such labeled data.

In our case the use of machine learning has a more complex architecture, but is shown in a simplified form in Fig. 1. We assume an ATN system described by distributed agents' behavior. ATN system behavior, or shortly ATN throughput, has many parameters and acts within an environment i.e. ATN throughput responds to environmental requirements. These environmental requirements can change with time, therefore they are referred to as Environment Behavior. The ATN system performance (throughput) should be measured with respect to Environment Behavior. Therefore, as shown in Fig. 1, our training data includes Environment Behavior, ATN throughput and some measures of resulting performance. In case of Empty Vehicle Management the best measures of the performance are related to throughput. The data describing ATN throughput can be collected during actual ATN runs and/or by simulations.

Different types of approaches can be used to determine ATN throughput. The typical approach is to develop a general evaluation functions, e.g. number of passengers transported per hour, the total distance for all passengers transported per hour, the average passenger waiting time, minimum energy, etc. These functions have a tendency to emphasize group interests rather than the individuals.

As in typical ML approach several modes are defined. In *training* mode, many working condition parameters (Environment Behavior) and many ATN system parameter sets are generated to obtain best ATN throughput tuning for the given Environment Behavior. Then, during testing mode and normal *execution* mode of the system, ML model will compute the parameters that will be used to control the individual vehicles as is shown in Fig. 2.

There are also possible improvement for this approach by applying additionally reinforcement learning. The architecture of such system will be very similar to Fig. 2, but with the additional arrow indicating that the throughput will be provided for ML system to describe the feedback to change ML model slightly by learning how to fine-tune parameters and to better adopt it to changing operation conditions.

## 2. ML USING NEURAL NETWORKS

We considered use of various machine learning (ML) algorithms including neural networks [26] [27] [28] [29] [30] [31], Support Vector Machines (SVM) [23], and random forest [32], to combine measurable features into a single number estimate of ATN throughput [15] [16] .

Neural networks were selected as ML solution since they support both traditional ML using feature vectors and Deep Learning using raw data. A neuron is a cell in the brain that is responsible for collecting processing and dissemination of electrical signals [24]. The brain's information processing capability is associated with the network of such neurons [33]. This is one of the reason that artificial neural network always were important part of Artificial Intelligence area and were of great interest to students.

Each neuron has n inputs with activation signals $x_0 \div x_n$. For each input a weight is assigned $w_0 \div w_n$. Positive value of the weight corresponds to activation and negative value of the weight deactivate. Output signal $y$, i.e. activation of the neuron, can be described as:

$$y = f(\sum_{i=0}^{n} x_i w_i) \quad (1)$$

Where function $f()$ – is an activation function. The activation function can be linear:

$$f(s) = as \quad for \quad a \in R \quad (2)$$

Very often the step functions are used as activation function or in case of more complex tasks the function of hyperbolic tangent or sigmoidal functions are applied.

Neurons can be combined into layers. There can be any number of neurons in each layer. Neural network is composed from many layers [34]. There is an input layer, the output layer and possibly a hidden layer. There can be any number of hidden layers within a neural network but one is usually enough to support typical problems.

The design of the neural network requires specification of number of neuron in each layer and the learning algorithm which is associated with the activation function. Let us discuss the design a neural net that will recognize data represented by Environment Behavior parameters.

Clustering Environment Behavior values and assuming small randomly distributed changes around the basic cluster can help to simplify ML training. Absence or presence of clustering of Environment Behavior defines different types of Neural Network. The first case is to use raw Environment Behavior data and to train ML models on such data. It means that the neural network has to accept the state of Environment Behavior vector as an input. It would result in the neural network number of neuron equal to 11 in the input layer [23]. The second NN would use values that label the clusters. That would make much easier to train ML models, but they might fail to pick best ATN throughput, especially when applied to Environment Behavior that is not clearly within a cluster. Our experiments suggested using richer environment data. Although this method requires larger training set, it provides a more precise response to the environment variable. Yet another NN can be constructed to improve decoding short term and long term tendencies and cyclic changes of the Environment. Such a NN would require additional neurons in the input layer to accommodate additional variable describing explicitly typical dynamics of Environment Behavior.

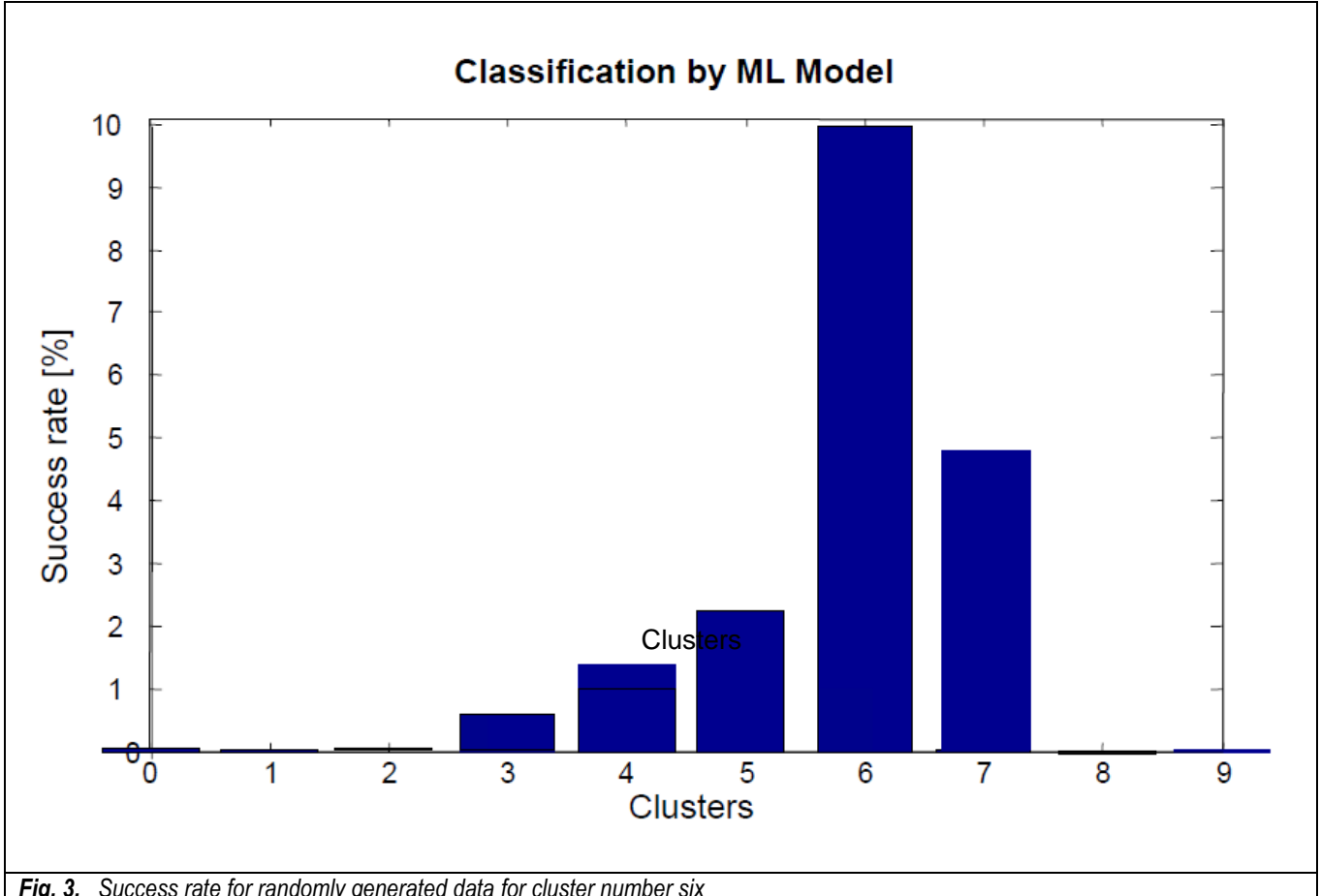

*Fig. 3. Success rate for randomly generated data for cluster number six*

There can be a similar discussion for clustering ATN System parameters values Absence or presence of clustering of ATN System parameters defines again different types of Neural Network. The first case is to use raw ATN system parameters and to train ML models to predict such data. The second NN would use values that label the clusters. That would make much easier to train ML models. Our experiments here, however, did not suggested using richer control space because of instances of overfitting. In order to provide a resilience to changes of Environment Behavior the cluster approach was preferred.

In our neural networks output need to be related to ATN Behavior parameters or their clusters. For example when we have 10 clusters then we might need 10 classifying states resulting in 10 neurons in the output layer [32]. Each neuron will have output either a 1 or zero. The value 1 indicates that it neural network recognized the given category. This happens if there is 100% of success in recognition of the ATN parameters. In the ambiguous situation that success rate is much lower, especially when the noise is present. The number of neurons in the hidden layer needs typically to be identified through experiments.

Once the neural network has been designed it needs to be trained. We mostly used supervised learning as shown in Figs. 1 and 2, which means that we feed the neural network with data set called a training set. After that the testing takes place that tries to identify an error related to applying the trained neural network to a data set called a testing set. Since typically the data for training set and testing set is obtained randomly from the provided set, the additional evaluation of the network is necessary. Thus each application requires extensive study to guarantee the proper performance of neural network.

Let us discuss the specific implementation of NN for ATN system. An important part of our efforts is to set up the proper environment consisting on appropriate ML library. One possibility is NN Toolbox [26]. We used another popular environment based on *theano*. Our environment and libraries included:

1. *theano* - provides fast multi-dimensional array calculation, allows GPU usage
    Code: import *theano*
2. *pandas* - converts csv files to DataFrames used for training.
    Code: import *pandas* as *pd*
3. *numpy* - numerical library for python
    Code: import *numpy* as *np*
4. *Keras* - Deep learning library built on top of Theano that allows for fast architecture iteration. Sequential - an object that encompasses deep learning operations
    Code: from *keras.models* import *Sequential*
5. *Add-ons*: Dense - provides access to the classic neural network layer. Dropout - gives access to the ability to randomly drop certain neurons during test time, which has been shown to dramatically improve performance. Activation - gives access to varying types of neural activators, for example relu and tanh
    Code: from *keras.layers.core* import *Dense, Dropout, Activation*
6. *Optimizers* - give access to different types of gradient descent algorithms for model training
    Code: from *keras.optimizers* import *SGD, Adam, RMSprop*
7. *Keras* utilities - give access to evalutation metrics
    Code: from *keras.utils* import *np_utils*
8. *sklearn* preprocessing - gives access to min max scaler
    Code: from *sklearn* import *preprocessing*
9. Custom library built for custom metrics
    Code: import *Throughput* as *Th*

The example execution of the ML model is shown in Fig. 3. The visualization tools displays discovered association of a given Environment Behavior with one of the 10 clusters constructed for ATN System parameters. The specific noise level was applied that modified Environment Behavior. More study are needed to formally determine the success rate for other clusters and for other space reduction methods with various controlled noise level.

We investigate the two algorithms: calling the vehicles for waiting passengers and empty vehicles redistribution. They both must be active in ATN operation for its best throughput. A quality measure for every algorithm is passenger waiting time, therefore every algorithm should be tuned separately, with fixed set of parameters of the other one. This procedure should be repeated for several times, switching the tuning between the two algorithms.

## 3. CONCLUSIONS

Our main focus was to facilitate "intelligent" solutions to be applied for ATN systems. We showed how we can include in ATN systems soft computing to produce simple but interesting "intelligent" behaviors. There are many benefits of using our testbed system with NN as the tool for improving performance of ATN systems including optimization of throughput as emphasized in this paper. More study are needed to compare ML results using raw data and various space reduction algorithms including clustering. Especially important is to formally determine the resilience of ATN systems to unexpectedly changing Environment Behavior.

So far, we have been exploring distributed protocols using simulation methods, with parameter sets assigned arbitrary. This is our first article in which we describe our experience of using ML for ATN. We considered several classification algorithms but we chose neural networks mainly for practical reasons because of the library *keras* which not only allows for "feature vector based" ML but also "Deep Learning". With the production of more simulation data we plan to use the same tool (keras) for "deeper" analysis reducing the role of "feature vector".

**Zastosowanie Uczenia Maszynowego (Machine Learning) do poprawy jakości działania zautomatyzowanych sieci transportowych (Automatic Transit Network-PRT)**

*W artykule omówiono nowe techniki usprawniania zautomatyzowanych sieci transportowych (ATN, wcześniej nazywanych Personal Rapid Transit - PRT), opartych na narzędziach sztucznej inteligencji. Głównym kierunkiem jest poprawa współpracy autonomicznych modułów, które używają protokołów negocjacyjnych w paradygmacie IoT. Jednym z celów jest zwiększenie przepustowości systemu transportowego poprzez dostrajanie współpracy autonomicznych pojazdów. Uczenie maszynowe (ML) jest wykorzystywane do poprawy algorytmów opracowanych przez programistów. Użyliśmy "istniejącego sterowania", odpowiadającego suboptymalnym rozwiązaniom, i skonstruowaliśmy modele dostrajania, aby dokładniej odnieść dynamikę systemu do jego wydajności. Mechanizm, który wpływa głównie na wydajność ATN to Zarządzanie Pustymi Pojazdami (Empty Vehicle Management - EVM). Wykorzystano algorytmy opracowane przez programistów: wzywanie pustych pojazdów dla oczekujących pasażerów i równoważenie w oparciu o realokację pustych pojazdów w celu osiągnięcia lepszej regularności ich rozmieszczenia. W tym artykule omówimy, jak można poprawić te algorytmy (i dostroić je do aktualnych warunków), używając ML do dostosowania indywidualnych zasad behawioralnych. Wykorzystanie technik ML było możliwe, ponieważ nasz algorytm oparty jest na zbiorze parametrów. Zestaw współczynników i progów może zostać dostrojony do podejmowania lepszych decyzji o planowaniu ruchu pustych pojazdów na torze.*

Authors:
**Bogdan Czejdo**, PhD – Department of Mathematics and Computer Science, Fayetteville State University, Fayetteville, NC 28301, USA, bczejdo@uncfsu.edu
**Wiktor B. Daszczuk**, PhD – Warsaw University of Technology, Institute of Computer Science, Nowowiejska str. 15/19, 00-665 Warsaw, Poland, wbd@ii.pw.edu.pl
**Mikołaj Baszun**, PhD – Warsaw University of Technology, Institute of Microelectronics and Optoelectronics, Nowowiejska str. 15/19, 00-665 Warsaw, Poland, M.Baszun@imio.pw.edu.pl